\RequirePackage[2020-02-02]{latexrelease}
\documentclass[preprint,showpacs,showkeys,preprintnumbers]{revtex4}

\usepackage{graphicx}
\usepackage{dcolumn}
\usepackage{bm}
\usepackage{epsfig}
\usepackage{epstopdf}
\usepackage{grffile}
\usepackage{color}
\usepackage{colordvi}
\usepackage{amssymb}
\usepackage{rotating}
\usepackage{lscape}
\usepackage{float}
\usepackage{mathtools}

\usepackage{footnote}
 \makesavenoteenv{environmentname}

\usepackage{hyperref}

\usepackage[T1]{fontenc} 

\begin{document}
\title{Zh Production, a Tool to Constrain 2HDM}

\author{Zeynab Bozorgtabar}
\email{bozorgtabarzeynab@stu.yazd.ac.ir}
\affiliation{Department of Physics, Yazd University, P.O. Box 89195-741, Yazd, Iran}

\author{Saeid Paktinat Mehdiabadi}
\email{spaktinat@yazd.ac.ir}
\affiliation{Department of Physics, Yazd University, P.O. Box 89195-741, Yazd, Iran}

\date{\today}

\begin{abstract}
Since the SM Higgs boson, h, is the superposition of two unequal-mass neutral Higgs bosons in 2HDM, therefore the associated production cross section for h and Z in SM and 2HDM can be different. In this paper, the possibility of using this difference to discover or constrain 2HDM is studied.
As a main result, the allowed parameter space of  2HDM  type-II is found in the light of  possible precise measurements from the LHC experiments. It is shown that combination of the constraints from these measurements with the previous constraints on 2HDM, from direct search and flavour physics, can rule out the main part of the 2HDM parameter space. 

\end{abstract}
\keywords{collider phenomenology, exclusion limits, production cross section, 2HDM, flavour physics}
\pacs{13.90.+i}
\maketitle

\section{Introduction}\label{sec:intro} 
The Standard Model of elementary particles (SM) describes the subatomic processes very precisely. During the last few decades SM has passed different tests to predict or explain  the experimental results. The Higgs mechanism was the last missing part of SM, explaining the masses of bosons and fermions. The discovery of a Higgs boson close to 125 GeV/$c^2$ was announced at the CERN Large Hadron Collider (LHC) in July 2012~\cite{ATLAS:2012yve}\cite{CMS:2012qbp}. The Higgs boson discovery opens a new era in experimental, theoretical and phenomenological particle physics.
There are many motivations behind the extension of SM \cite{Lykken:2010mc}. From baryogenesis to neutrino mass and from hierarchy problem to dark matter are few examples of the shortcomings of SM.
One of the simplest extensions of the SM is called Two-Higgs-Doublet-Model (2HDM)~\cite{Ma_1994}~\cite{HABER198575}~\cite{Carena_2003}~\cite{PhysRevD.72.035004}~\cite{DJOUADI_2008}~\cite{Branco:2011iw}. As the name suggests, the idea is to add two Higgs doublets instead of one Higgs doublet in the standard Higgs mechanism. After spontaneous electroweak symmetry breaking, it ends up having 5 Higgs particles where two of them are CP-even like the already discovered SM Higgs boson. In the 2HDM point of view, the SM Higgs boson is the linear combination of this two unequal mass Higgs bosons. So, the associated production cross section of Z boson and h (Zh) can be different in SM and 2HDM. This paper studies the effect of this measurement on the 2HDM allowed parameter space. Already, a similar idea is used by the second author in Ref.~\cite{Ayazi:2019kli}. The analysis uses proton-proton collisions at $\sqrt{s}$ = 13 TeV.  Although there are some measurements of the Zh production cross section from the LHC experiments, however they do not have enough precision. So, there is a room for new physics and a suitable motivation for this research work. The most recent results can be found in Refs.~\cite{CMS:2023vzh} ~\cite{ATLAS:2023qpu}. For example, the best signal strength reported by the CMS collaboration is $\mu = 1.15 ^{+0.22} _{-0.20}$ and 
the best signal strength reported by the ATLAS collaboration is $\mu = 1.28_{-0.29}^{+0.30} (stat.) _{-0.21}^{+0.25} (syst.)$, which are still far from a precise measurement. 

In continue, direct search constraints are added up on top of the constraints from the cross section analysis. The allowed region from the combination of two different analysis is very limited. At the end, consistency of the favorite parameter space with the constraints from some observables of flavour physics is investigated.

To explain the analysis and report the results, the paper is organized as follows. A brief phenomenology of 2HDM is discussed in Sec.\ref{sec:pheno}. In Sec.\ref{sec:ana}, Zh production and its cross-section prediction is described. The results of the cross section analysis, direct search, and flavour physics are shown in Sec.\ref{sec:results}, Sec.\ref{sec:hb}, and Sec.\ref{sec:flavour}, respectively. Section \ref{sec:con} concludes the paper.

\section{Phenomenology of 2HDM}\label{sec:pheno} 
One of the simplest extensions of the Higgs sector of SM is to add a new doublet. It is the main idea of the two Higgs Doublet Model (2HDM). In this model two scalar Higgs doublets $\Phi_{i}$, $i$ = 1, 2 are introduced: 
\begin{equation}
	\Phi_{i}= \binom{\phi_{i}^{\dag}}{\frac{1}{\sqrt{2}}(v_{i}+\phi_{i}+i\chi_{i})}.
\end{equation}

Higgs doublets couple to fermions proportional to their masses in different types of the model. On the basis of, which type of fermions are coupled with which doublet, these models are divided into four different types. Each type of 2HDM features special constraints and phenomenologies~\cite{PhysRevD.74.015018}. In this analysis, the  2HDM type-II~\cite{Eberhardt:2013uba} is used for illustration, where all the up type quarks couple with the second Higgs doublet, $\Phi_{2}$, and all the down type quarks along with the charged leptons couple with the other Higgs doublet, $\Phi_{1}$.
There are fourteen parameters in very common scalar potential. The scalar potential of 2HDM, after the assumption of CP conservation (all real parameters), and absence of flavour-changing neutral currents (FCNC) at tree level (softly broken $Z_2$ symmetry) can be written by $\Phi_{1}$ and $\Phi_{2}$ as follows~\cite{Ginzburg_2005}:
\begin{eqnarray}
	V_{2HDM} &=& m_{11}^{2} \Phi_{1}^{\dagger}\Phi_{1}
	+ m_{22}^{2} \Phi_{2}^{\dagger}\Phi_{2}
	-   m_{12}^{2}(\Phi_{1}^{\dagger}\Phi_{2} 
	+ \Phi_{2}^{\dagger}\Phi_{1}) 
	+ \frac{1}{2}\lambda_{1}(\Phi_{1}^{\dagger}\Phi_{1})^{2} 
	+ \frac{1}{2}\lambda_{2}(\Phi_{2}^{\dagger}\Phi_{2})^{2}\label{eq.V}\\\nonumber
	&+& \lambda_{3}(\Phi_{1}^{\dagger}\Phi_{1}) (\Phi_{2}^{\dagger}\Phi_{2}) 
	+ \lambda_{4}(\Phi_{1}^{\dagger}\Phi_{2}) (\Phi_{2}^{\dagger}\Phi_{1}) + \frac{1}{2}\lambda_{5}((\Phi_{1}^{\dagger}\Phi_{2})^{2}+(\Phi_{2}^{\dagger}\Phi_{1})^{2}).
\end{eqnarray} 
In equation~\ref{eq.V}, $\lambda_{i}$, where $i = 1, 2...,5$, are coupling parameters having no dimensions and $m_{11}^{2}$ , $m_{22}^{2}$ and $m_{12}^{2}$ are squared mass parameters. 
When the electroweak symmetry breaks, each scalar doublet acquires a vacuum expectation value which must satisfy $ v_{1}^{2} + v_{2}^{2}= (246  GeV)^{2}\equiv v_{SM}^{2} $~\cite{PhysRevD.72.035004} and is given as \cite{Eberhardt:2013uba}:
\begin{equation}
	\langle\Phi_{1}\rangle = \frac{1}{\sqrt{2}} \binom{0}{v_{1}}  ,
	\langle\Phi_{2}\rangle = \frac{1}{\sqrt{2}}\binom{0}{v_{2}}. 
\end{equation}
There are eight degrees of freedom in 2HDM that after electroweak symmetry breaking three of those give mass to the gauge bosons, $W^\pm$ and $Z$. The remaining five states appear as physical scalars. Among them there are two neutral CP-even scalar (h, H), one neutral CP-odd pseudoscalar (A) and one pair of charged Higgs bosons (H$^\pm$).

In general, the SM Higgs boson would be a linear combination of the physical CP-even scalars i.e. h and H \cite{Branco:2011iw}:
\begin{equation}	
	h_{SM}  = H \cos(\beta-\alpha) + h \sin (\beta-\alpha) 
	\label{eq.Hh}
\end{equation}
Where, $ \alpha $ is the mixing angle of the neutral CP-even Higgs bosons  and the angle $\beta $  is defined by the ratio of the vacuum expectation values:
\begin{equation}
	\tan\beta = v_{2} /v_{1}
\end{equation}
In the alignment limit (sin$(\beta-\alpha)=1$), the lighter neutral CP-even Higgs boson (h) has exactly the same couplings as the SM Higgs boson. By getting distance from the alignment limit, the couplings of h in 2HDM can be very different from those of the SM Higgs. In this analysis, the effect of this difference in production of Zh in 2HDM and SM is studied in more details.

\section{Associated production of Z and Higgs bosons}\label{sec:ana} 
\begin{figure}[!htb]
	\includegraphics*[width=.5\textwidth]{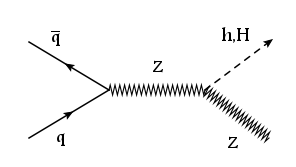}
	\caption{The main process to generate Zh is Higgs Strahlung.}
	\label{HiggsStrahlung}
\end{figure}

In this study, a comprehensive analysis on Zh, the associated production cross section of a SM vector boson, Z, and a neutral Higgs boson, h, in the 2HDM framework is reported. In the LHC experiments, two energetic incoming proton beams collide at a center of mass energy of $\sqrt{s}$ = 13 TeV. Figure~\ref{HiggsStrahlung}
shows a typical example of the events that produce a Zh final state. This diagram is responsible for more than 98\% of the Zh events. Although the radiation of a Higgs from the produced Z (Higgs Strahlung) is the most important source of the Zh production, in both SM and 2HDM, but all tree level sources of production are considered in this analysis. The SM prediction for the leading order cross section of this process is 0.5795 $\pm$ 0.0004 pb.  

After the discovery of the Higgs boson in 2012, some of the SM Higgs boson searches at LHC have been reappraised to search for a heavy CP-even state.
This state could be the heavy CP-even Higgs boson of a 2HDM, or a generic additional singlet. 
In both cases, the mass and the natural width of the additional H state can be very different from the corresponding parameters of the SM Higgs boson. If such a particle exists, Zh production cross section would not match the value of SM (See Eq.\ref{eq.Hh}). This difference is used to constrain the 2HDM parameter space.
To illustrate the idea, 2HDMC-1.8.0 \cite{Eriksson:2009ws} is used to calculate the 2HDM couplings and mass widths by giving the input parameters. The `physical basis' is used to define a test point. In this basis, 4 Higgs masses out of 5, $\tan \beta$, $m_{12}^2$, and $\sin (\beta - \alpha)$ are used as the input parameters. The produced  couplings/branching ratios are passed as a model to MadGraph-2.6.7 \cite{Alwall:2011uj} to generate 20k events. This statistic is sufficient to have a reasonable statistical uncertainty on the measured cross sections. Apart from the center of mass energy and 2HDM parameters, the rest of the variables are the default variables of MadGraph-2.6.7. The ROOT-5.34/30 analysis framework \cite{Brun:1997pa} is utilized to depict the results.

All 2HDM parameters in `physical basis' are examined to find the important variables affecting the production cross section of Zh in 2HDM. At the end only $\sin(\beta-\alpha)$ and the mass of the heavy CP-even Higgs ($M_H$) are found to be relevant. In the rest of the analysis, the other parameters are set as follows:  $\tan\beta = 5$, $M_A$ and ${M_{H^{\pm}}}$ are taken equal to 800 GeV/$c^2$, also $m_{12}^{2}$ value is set to 1000 GeV$^2/c^4$. In order to satisfy the theoretical constraints on 2HDM parameter space, e.g., tree-level unitarity, perturbativity and vacuum stability, $m_{12}^{2}$ can be changed, but since this parameter does not affect the results, it is fixed to 1000  GeV$^2/c^4$ without further tuning.

\section{Cross Section Constraints}\label{sec:results}
\begin{figure}[!htb]
	\includegraphics*[width=.49\textwidth]{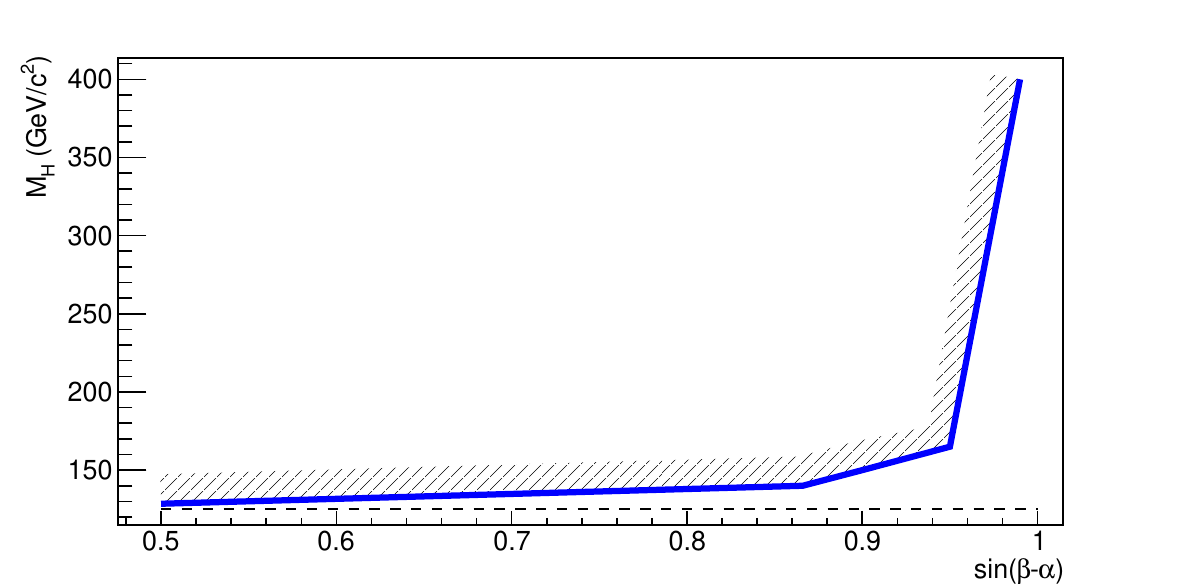}
	\includegraphics*[width=.49\textwidth]{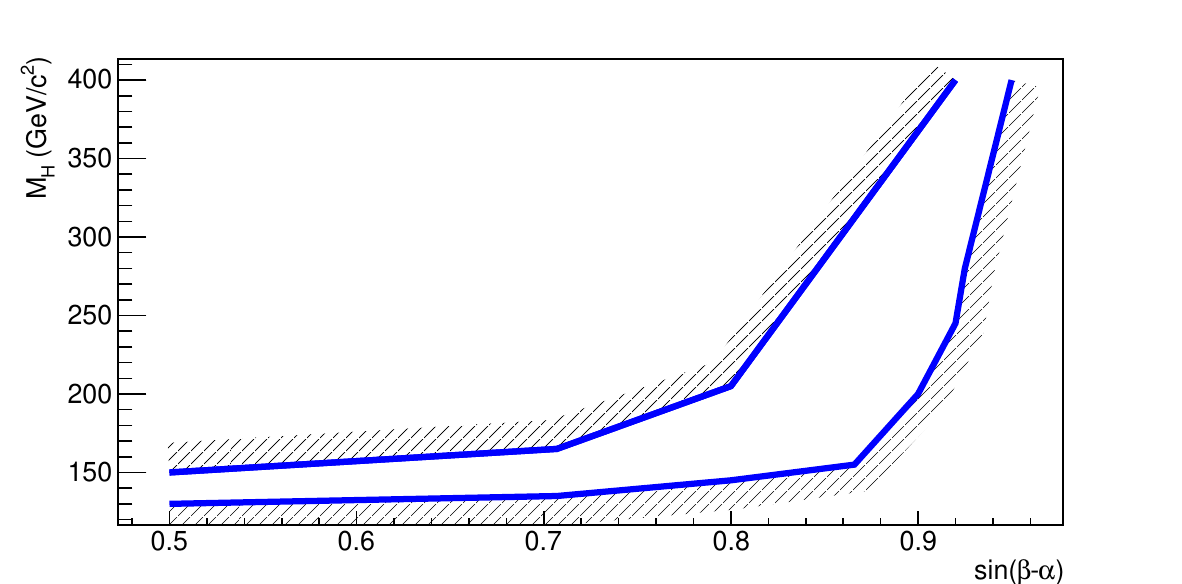}
	\caption{The excluded regions (shaded regions) in the relevant 2HDM parameter space when 100\%(left) or 75\%(right) of the SM cross section is observed.} 
	\label{fig:Ex}
\end{figure}

Since still there is not a precise experimental measurement of Zh production cross section available, two scenarios are considered. 
In the first scenario, the measured cross section at LHC is assumed to be same as the SM prediction with 10\% uncertainty. In the light of this measurement, the allowed region of 2HDM parameter space is found and shown in Fig.~\ref{fig:Ex}(left).
If the cross section of a point in the parameter space is within 1$\sigma$ of the SM prediction, the point is marked as allowed, otherwise it is excluded \cite{Bini:2014yhn}. 
It can be seen that only the points with $M_H$ close to 125 GeV/$c^2$ or the points with $\sin(\beta - \alpha)$ close to 1.0 are allowed and the rest of the parameter space is excluded. For the former points, h and H are degenerate in mass, so there is not any difference between SM and 2HDM in this part. But, this degeneracy can breaks down the perturbativity of the model and is not favoured by the theoretical constraints \cite{KANEMURA1993155} \cite{Gunion:1989we} \cite{Branco:2011iw}. For the latter points, according to Eq.\ref{eq.Hh}, the SM Higgs boson is purely made of h and it does not receive any contribution from H, so independent of $M_H$, all points with $\sin(\beta - \alpha)$ close to 1.0 are allowed.

In the second scenario, it is assumed that the measured cross section at LHC is equal to $75\%$ of the SM predicted cross section. Its uncertainty is again assumed to be 10\%. Figure \ref{fig:Ex}(right) shows the excluded regions in this scenario. The reduction in cross section is coming from either the heavier H (horizontal band) or less h component in the SM Higgs boson (vertical band). The latter case can be understood from Eq.\ref{eq.Hh}.

The cross section of a point in the parameter space is sum of the production cross section of  Zh and ZH final states. A realistic analysis needs to simulate the detector effects and apply selection cuts on the invariant mass of the products of the h and H decays and separate the two final states but, it is ignored here for simplicity.
It should also be emphasized that the cross section values in this analysis are in the leading order approximation and the results could change if higher order corrections~\cite{Degrande_2015}~\cite{Denner_2018} are considered, but here the idea is only to proof the principle of the method.
Due to the strong constraints that can be introduced by this analysis, the LHC experiments are encouraged to increase the precision of the Zh production cross section in near future to either discover or exclude 2HDM. 

\section{Direct Search Constraints}\label{sec:hb}
To take into account the results of the previous experimental searches for 2HDM Higgses, the HiggsBounds-5.7.0 (HB) code \cite{Bechtle:2020pkv} is used. 
This code checks the consistency of the tested point of the 2HDM parameter space with the current experimental constraints and decides if the point is excluded or allowed. 

\begin{figure}[!htb]
	\includegraphics*[width=.49\textwidth]{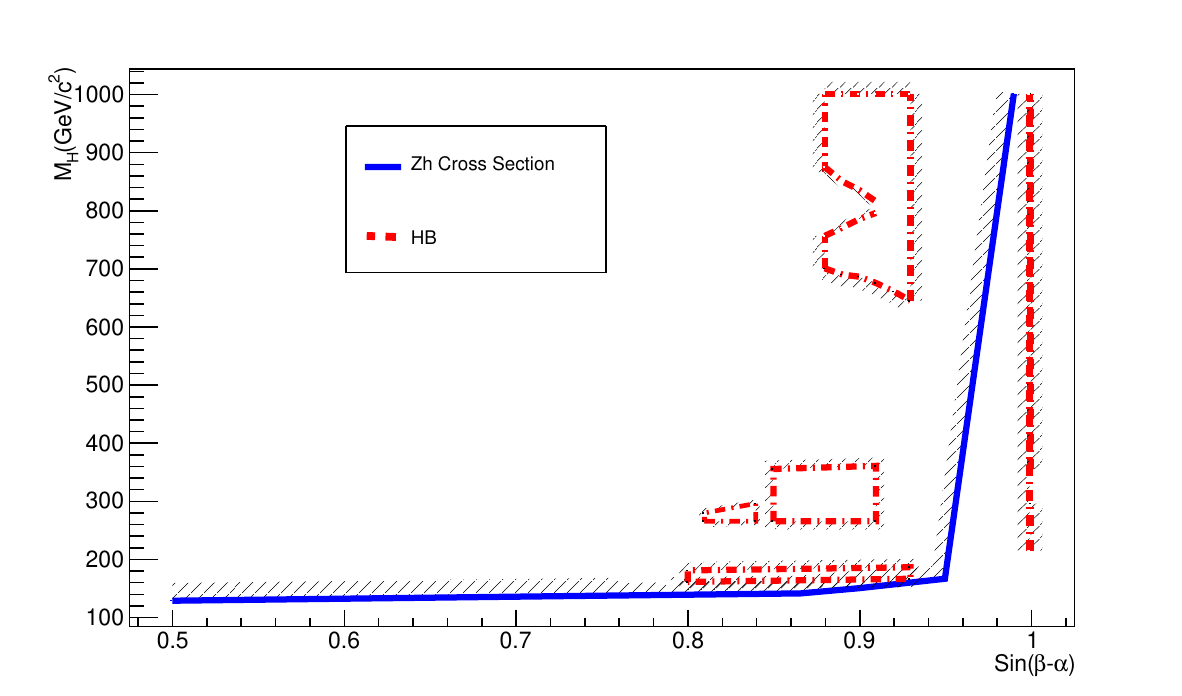}
	\includegraphics*[width=.49\textwidth]{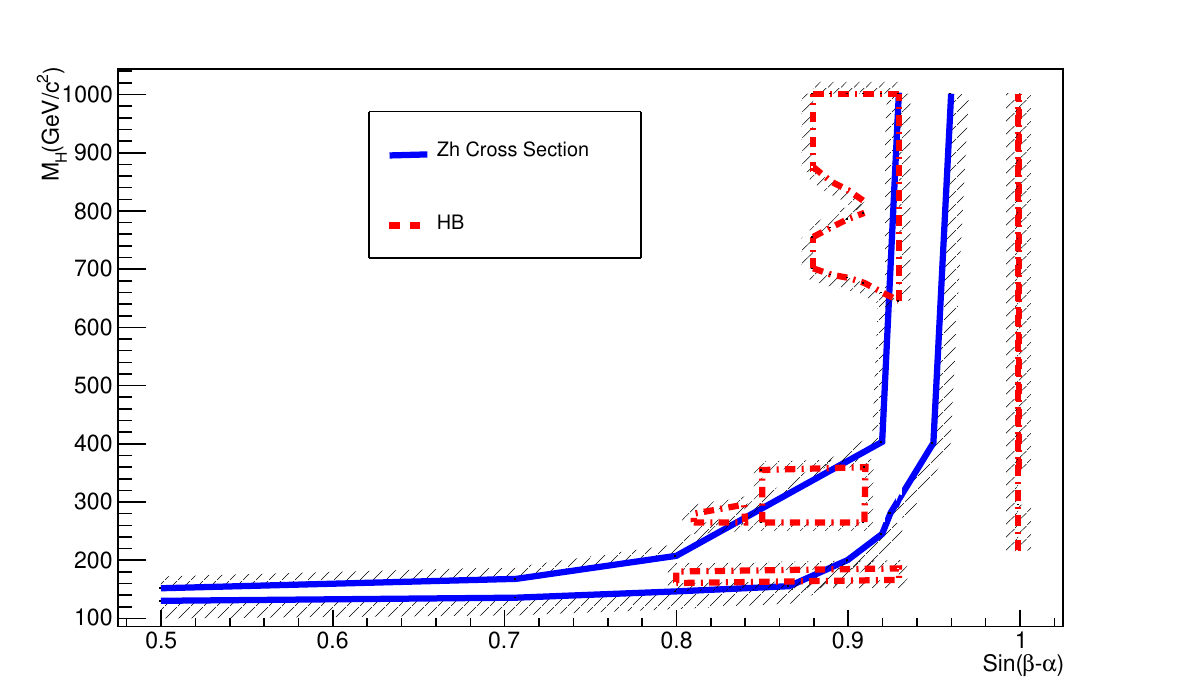}
	\caption{The allowed and excluded regions in the relevant 2HDM parameter space after using HB.}
	\label{fig:ExHB}
\end{figure}

The results are shown as two colored lines in Fig. \ref{fig:ExHB}. 
The red color lines show the result of HB and the blue color line shows the result of this analysis. The left and right plots show the results of the first and second scenario, respectively. It can be seen that the allowed regions after combination of the results are very small. 
In each case, the left side of figures are completely excluded. The used parameters to find the HB results are sames as the parameters of the main analysis. 

\begin{figure}[!htb]
	\includegraphics*[width=.49\textwidth]{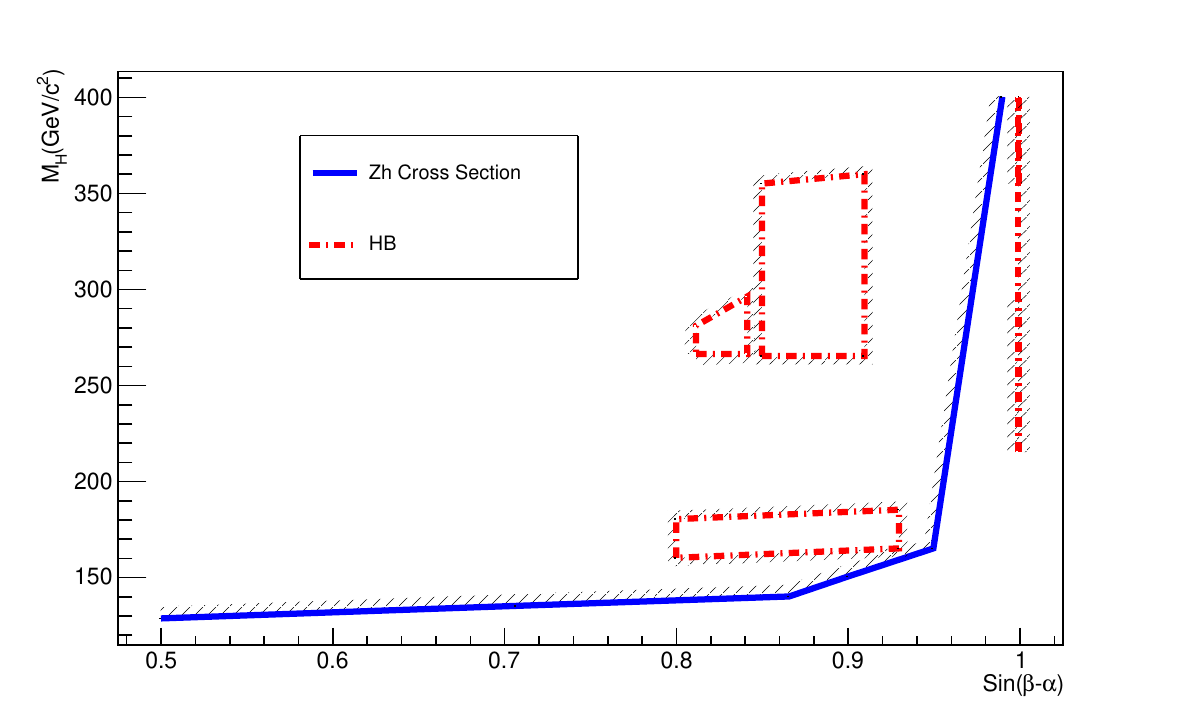}
	\includegraphics*[width=.49\textwidth]{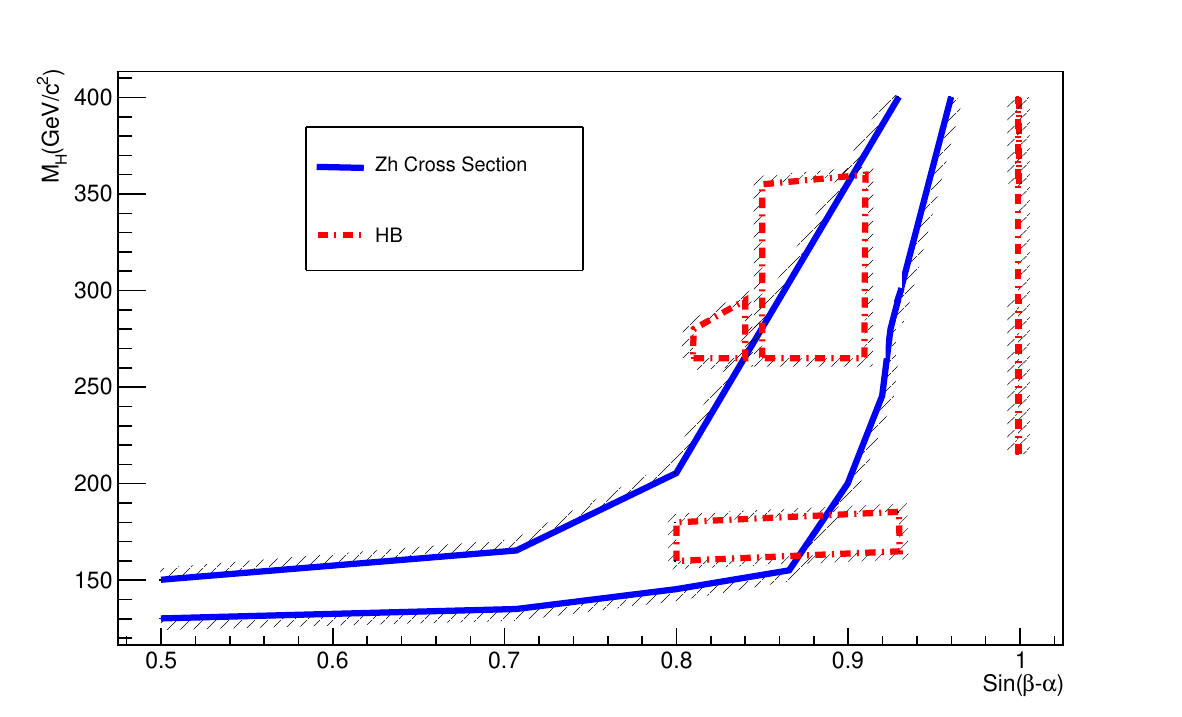}
	\caption{The allowed and excluded regions in the 2HDM parameter space after using HB for low $M_H$.}
	\label{fig:ExHBZoomed}
\end{figure}

To have a better view of the results, the H masses under 400 GeV/$c^2$ are shown separately in Fig. \ref{fig:ExHBZoomed}. 
It can be seen that for the first scenario, only some regions close to the alignment limit with $M_H$  above 220 GeV/$c^2$ are allowed. In the second scenario, the alignment limit is completely excluded and only some regions with $\sin(\beta - \alpha)$ between 0.8 and 0.87 with $M_H$ between 150 and 170 GeV/$c^2$ and some parts of the rectangular, made by  $\sin(\beta - \alpha)$ between 0.85 and 0.9 and $M_H$  between 260 and 350 GeV/$c^2$, are allowed.

\section{Flavour Physics Constraints}{\label{sec:flavour}}
As described in the introduction section, despite the tremendous and enormous successes of SM, this model is considered as an effective low energy description of nature and there are several reasons that it should be extended. The main part of these extensions suggest some new particles. One way to search for new particles is to directly produce them in as-yet unreachable energies. Another way to search for new physics signals is to look for indirect signs of unknown heavy particles. The flavour physics can be used to study such signatures.

In particle physics, flavour is a property to distinguish the different particles of two groups of basic fermions, i.e. quarks and leptons. There are six flavours of quarks and six flavours of leptons. In the other words,  
when the phenomenology of processes depend on the flavour quantum numbers, it belongs to flavour physics. The flavour is commonly used as the name of 
the particle given to different versions of the same type of particle. 
The flavour physics is usually used to detect a signal of new physics which is a rare process in SM and can only happen in higher orders of radiative corrections but, can receive a non-negligible contribution from new physics. 
So, the observation of such processes is interpreted as a signal of the existence of new physics \cite{Bifani:2018zmi}, \cite{ParticleDataGroup:2022pth}.
The charged current weak force is the only interaction that can change the flavour.

\subsubsection{B-mesons}
One of the main parts in flavour physics is B-physics that depends on processes relevant to B hadrons. These hadrons consist of at least one b-flavoured quark. Depending on type of the constituent quarks, they can be charged or neutral. One of the neutral types of B-meson is the $B_{s}^{0}$ particle. This particle has an anti-bottom quark and a strange quark ($\bar{b}s$). A possible decay mode of the  $ B_{s}^{0}$ particle is to a muon-antimuon pair: $B_{s}^{0} \rightarrow \mu^{+} \mu^{-}$.

\begin{figure}[!htb]
	\includegraphics*[width=.49\textwidth]{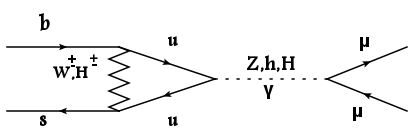}
	\includegraphics*[width=.49\textwidth]{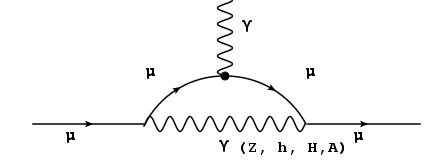}
	\caption {A sample of Feynman diagrams contributing to the BR($ B_{s}^{0}\rightarrow \mu^{+}\mu^{-} $) and $ \delta a_{\mu}. $}
	\label{fig:flavour}
\end{figure}

The FCNC leptonic decay of this meson ($B_{s}^{0} \rightarrow \mu^{+} \mu^{-}$) is especially sensitive to the contributions of scalar operator in new physics models. Therefore, it can be an excellent probe of the 2HDM effects and provides a good framework to search for new sources of FCNC decays. The decay mode is strongly suppressed within SM, because there is not any tree level FCNC contribution. Beside this, due to having at least one off-diagonal element of CKM matrix, its probability will decrease highly. The $B_{s}^{0}$ meson is a spin zero particle, so the decay is helicity suppressed also. The different types of suppression on this decay mode result in a very small branching ratio. However, it can obtain a larger branching ratio in special extensions of SM, like 2HDM. Figure ~\ref{fig:flavour}(left)
shows the schematic contribution of 2HDM Higgs bosons in dileptonic decay of $B_{s}^{0}$ meson. 

The above discussion is applicable for $B_{d}^{0}$ also. In parton model, this meson is the bound state of $\bar{b}d$. 

\subsubsection{D-mesons}
Hadrons containing a charm quark, make another class of observables which are used in flavour physics. In this analysis, some of the D-meson observables are used to find the preferred parts of the parameter space. Depending on the charge of a D-meson its constituents are a charm quark and a quark from the first family of quarks (up and down quarks). If the quark next to the charm quark is a strange quark, the meson is called $D_s^{\pm}$. The branching ratio of muonic decay of the charged D-mesons, BR($D^{\pm}\rightarrow \mu^{\pm}\nu$) and BR($D_s^{\pm}\rightarrow \mu^{\pm}\nu$), are studied here. It is obvious from what was discussed above that these branching ratios can be affected by the existence of the 2HDM particles and measuring them can constrain the 2HDM parameter space.

\subsubsection{The Muon's Anomalous Magnetic Moment}
Another sensitive observable which is investigated in this analysis is the muon's anomalous magnetic moment, $a_{\mu}$. 

The $ a_{\mu} $ measures the deviation of the muon gyromagnetic moment $ g_{\mu} $ from 2 due to quantum loop effects:
\begin{equation}
	a_{\mu} =\frac{(g_{\mu} - 2)}{2} 
\end{equation}
The deviation in $a_{\mu}$ can be a tool to find the new physics effects such as 2HDM contributions. Since 2017, measurements show a significant deviation from the SM prediction for this observable. The latest measurements show a 4.2$\sigma$ deviation, which is no more negligible ~\cite{ParticleDataGroup:2022pth}.
This deviation is called  $ \delta a_{\mu} $
\begin{equation}
	\delta a_{\mu} = a_{\mu}^{exp} - a_{\mu}^{SM}
\end{equation}
where in the subtraction, the former value is the experimental measurement and the latter one is the SM prediction. Figure ~\ref{fig:flavour}(right) shows how the 2HDM particles can enter and modify the Feynman diagrams related to $g_{\mu}$ calculation.

\subsubsection{SuperIso}
SuperIso v4.1 \cite{Neshatpour:2022fak}, \cite{Mahmoudi:2008tp} is a computational tool to investigate the effect of new physics on flavour physics observables. Different parameters and masses defining a test point in 2HDM parameter space are passed to SuperIso to calculate the flavour physics observables. The calculated values are compared with the measured values of these observables to decide if the test point is consistent in one or two $\sigma$ precision. 
The measured values and the allowed intervals for each observable  \cite{ParticleDataGroup:2022pth} are shown in table \ref{tab:Values}.
\begin{table}
	\caption{The measured values of the B-mesons  (D-mesons) observables and  $\delta a_{\mu}$ multiplied by  $10^{9}$ ($10^{4}$)\cite{ParticleDataGroup:2022pth}.}
	\label{tab:Values}
	\begin{tabular}{|c|c|c|c|c|c|}
		\hline
		Observable  & BR($ B^0_{s}\rightarrow \mu^{+} \mu^{-}$) & BR($ B^0_{d}\rightarrow \mu^{+} \mu^{-}) $ &  $\delta a_{\mu}$ &BR($D^{\pm}\rightarrow \mu^{\pm}\nu$)& BR($D_s^{\pm}\rightarrow \mu^{\pm}\nu$)\\ [1ex]
		\hline
		Value &  3.01$\pm$0.35 & $ 0.07 _{-0.11}^{+0.13} $ &  2.51 $\pm$ 0.59 &3.74 $\pm$ 0.17 &5.43$\pm$0.15\\\hline
		$ 1\sigma $ Interval &   2.66 to 3.36 &  -0.04 to 0.20 & 1.92 to 3.10& 3.57 to 3.91 & 5.28 to 5.58\\\hline
		$ 2\sigma $ Interval &    2.31 to 3.71 &  -0.15 to 0.33 &  1.33 to 3.69&3.40 to 4.08& 5.13 to 5.73\\
		\hline
	\end{tabular}		
\end{table}

Several flavour observables are implemented in SuperIso. Here only 
BR($B_{s}^{0} \rightarrow \mu^{+} \mu^{-}$), 
BR($B_{d}^{0} \rightarrow \mu^{+} \mu^{-}$),
BR($D^{\pm}\rightarrow \mu^{\pm}\nu$), 
BR($D_s^{\pm}\rightarrow \mu^{\pm}\nu$), and
$\delta a_{\mu}$ are considered.  The former one is referred to as BRBs in continue. It was seen in Sec.\ref{sec:results} 
that for Zh cross section analysis, only the mass of the heavy CP-even Higgs plays a role, so only the flavour observables than can receive contribution from the heavy CP-even Higgs boson are studied here. 

Different 2HDM parameters are tested to find their effect on the favorite flavour observables. At the end only five independent parameters, $ M_{A} $, $ M_{H} $, $ M_{H^{\pm}} $, $\tan\beta$ and $\sin(\beta-\alpha)$, are found as the parameters affecting the observables. Among these set of input parameters, three of them are found more important: the mass of the charged Higgs boson, $ M_{H^{\pm}} $,  $\sin(\beta-\alpha)$, and $ \tan\beta $. The other two parameters ($ M_{A} $ and $ M_{H} $), although change the observable values but, the changes are very minor.
In consistency with the previous sections, in continue, different 2HDM parameters are set as follows: $ M_{h} = 125 $ GeV/$ c^{2} $, $ m_{12}^{2} =1000 $ GeV$^2/c^4$. The values of other parameters are indicated in the following figures or their descriptions.

\begin{figure}[!htb]
	\includegraphics*[width=.65\textwidth]{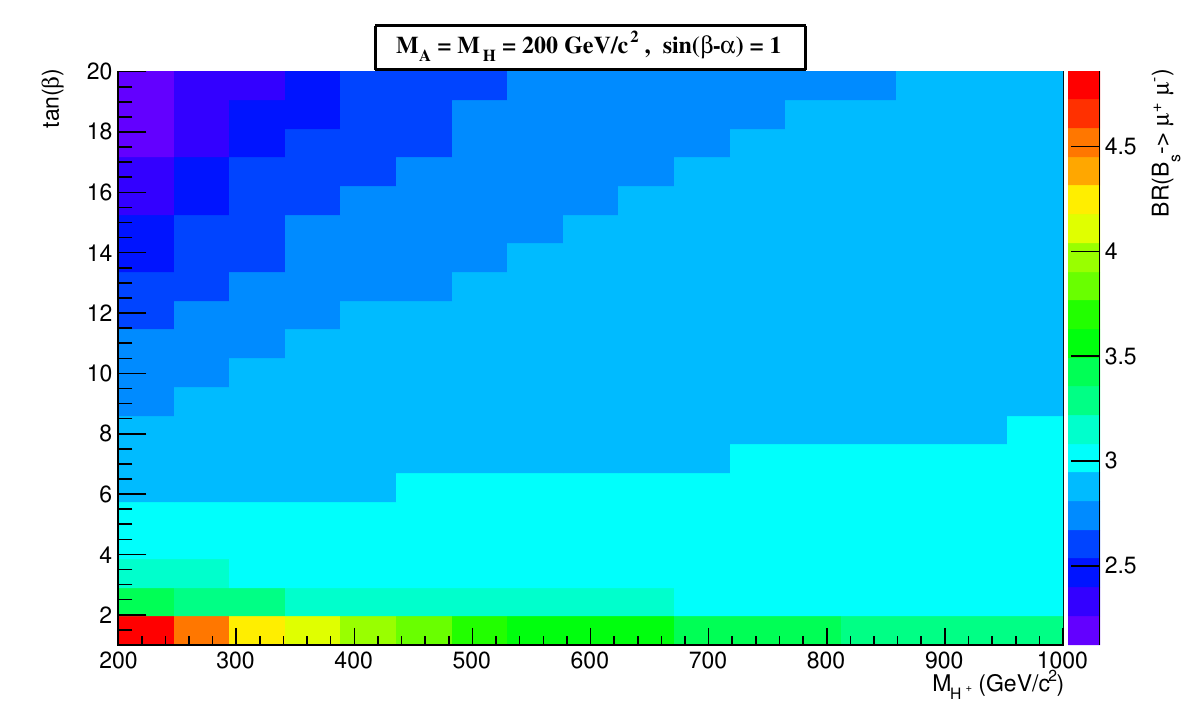}
	\caption{The BRBs values in ($M_{H^{\pm}}$, $\tan\beta$) plane.}
	\label{fig:Bs1000}
\end{figure}

\begin{figure}[!htb]
	\includegraphics*[width=0.65\textwidth]{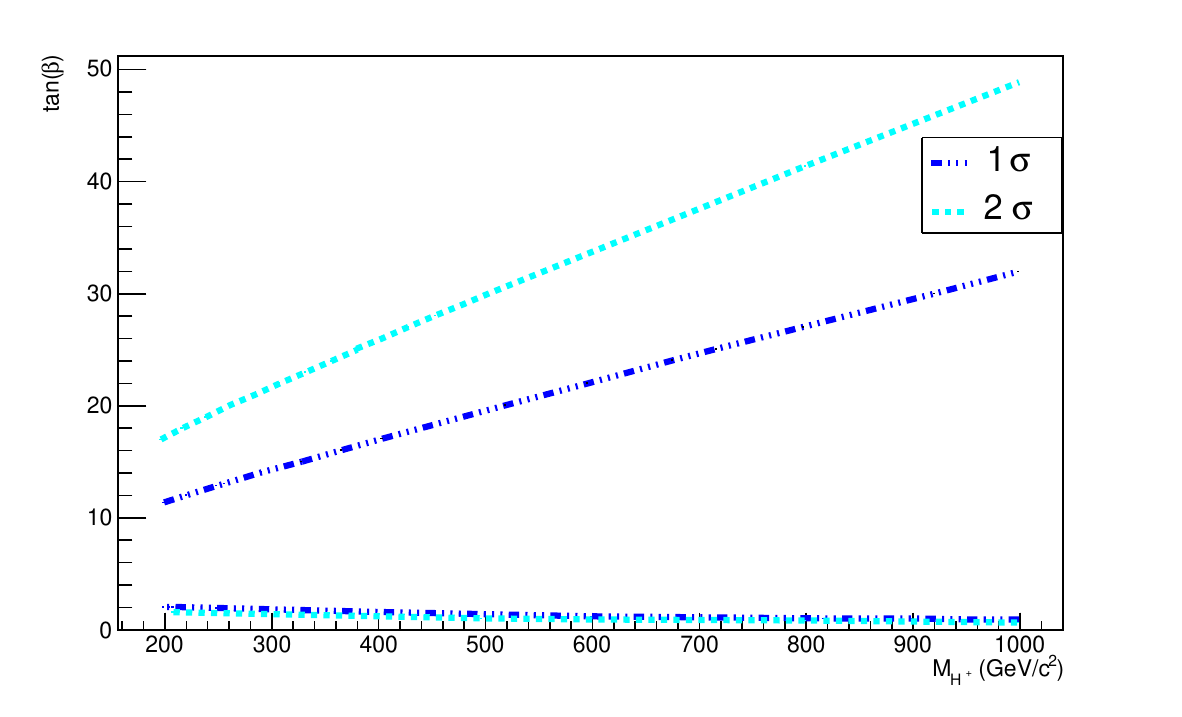}
	\caption{The regions consistent with BRBs measurement at $ 1\sigma $ and $ 2\sigma $ precision.}
	\label{fig:precision}
\end{figure}

The BRBs values in ($M_{H^{\pm}}$, $\tan\beta$) plane is shown in  Fig.\ref{fig:Bs1000}.
The CP-odd Higgs (A) and the heavy CP-even Higgs (H) are degenerate and have masses equal to 200 GeV/$c^2$. The distribution is shown in alignment limit but, it is found that BRBs depends on $\sin(\beta-\alpha)$ very weakly.
Figure ~\ref{fig:precision}
illustrates the boundaries of consistent regions with BRBs measurement in $1\sigma$ and $2\sigma$ precision. The regions in  $\tan\beta <2$  are not consistent for all masses. However, in high $\tan\beta$ depending on the charged Higgs mass the upper bound of this parameter can change between 10 to 30, in $1\sigma$ precision.

\begin{figure}[!htb]
	\includegraphics*[width=.65\textwidth]{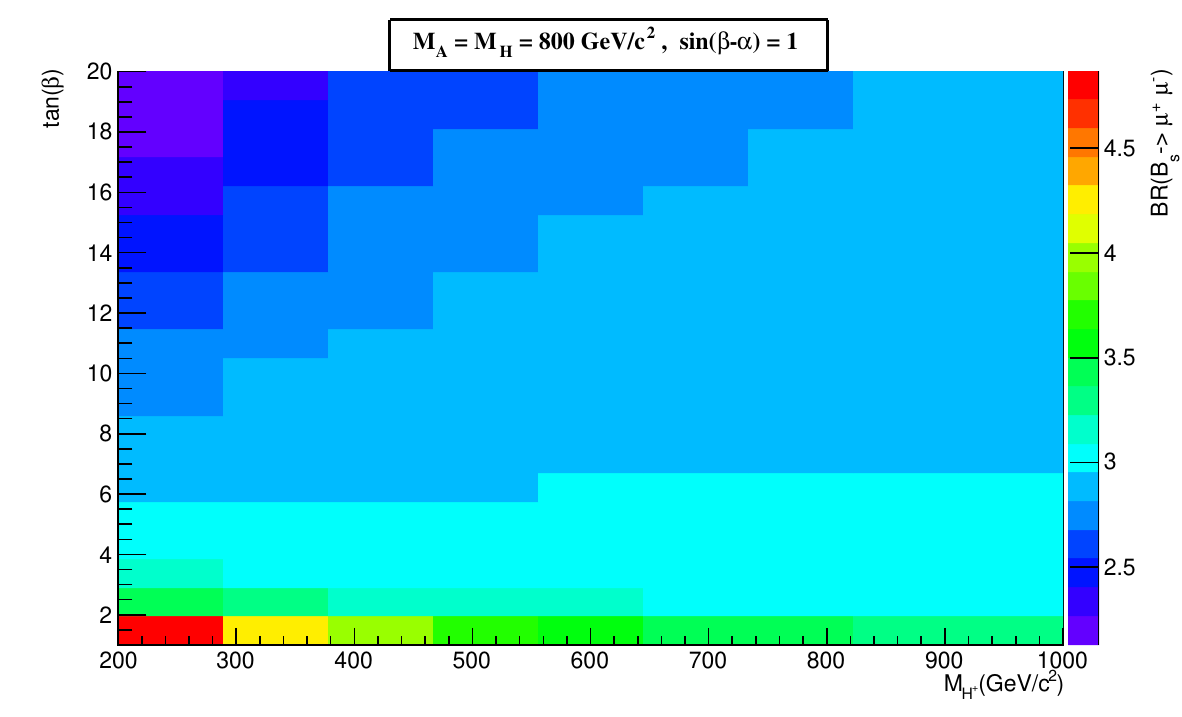}
	\caption {Same as Fig.\ref{fig:Bs1000}, when other Higgs masses are changed.}
	\label{fig:Bs1000A800}
\end{figure}

Similar distribution of Fig.\ref{fig:Bs1000} is shown in Fig.\ref{fig:Bs1000A800}, 
when  $M_{H}$ = $M_{A}$ = 800 GeV/$c^2$. It can be seen that the changes are minor.

It is found that the constraints caused by the measurements of BR($B_{d}^{0} \rightarrow \mu^{+} \mu^{-}$) are weaker than the constraints from BRBs ones. Investigating that observable does not add anything special to this analysis, so the results are not shown here.

Studying the constraints due to $D^{\pm}$ and $D_s^{\pm}$ shows that their branching ratios are almost constant in the whole parameter space which is investigated in this analysis. Changing $M_{H^{\pm}}$, $M_{H}$, and $M_{A}$ from 200 to 1000 GeV/$c^2$, $\sin(\beta-\alpha)$ from 0.9 to 1, and $\tan\beta$ from 1 to 20 result in almost the same values for the branching ratios. The changes in the branching ratios due to changing the parameters is about few 0.001. The quantities of BR($D^{\pm}\rightarrow \mu^{\pm}\nu$) and BR($D_s^{\pm}\rightarrow \mu^{\pm}\nu$) are very close to 4.14 and 5.25, respectively. The former value lies out of $2\sigma$  interval but, the latter one lie between $1\sigma$ and $2\sigma$ interval.

\begin{figure}[!htb]
	\includegraphics*[width=.65\textwidth]{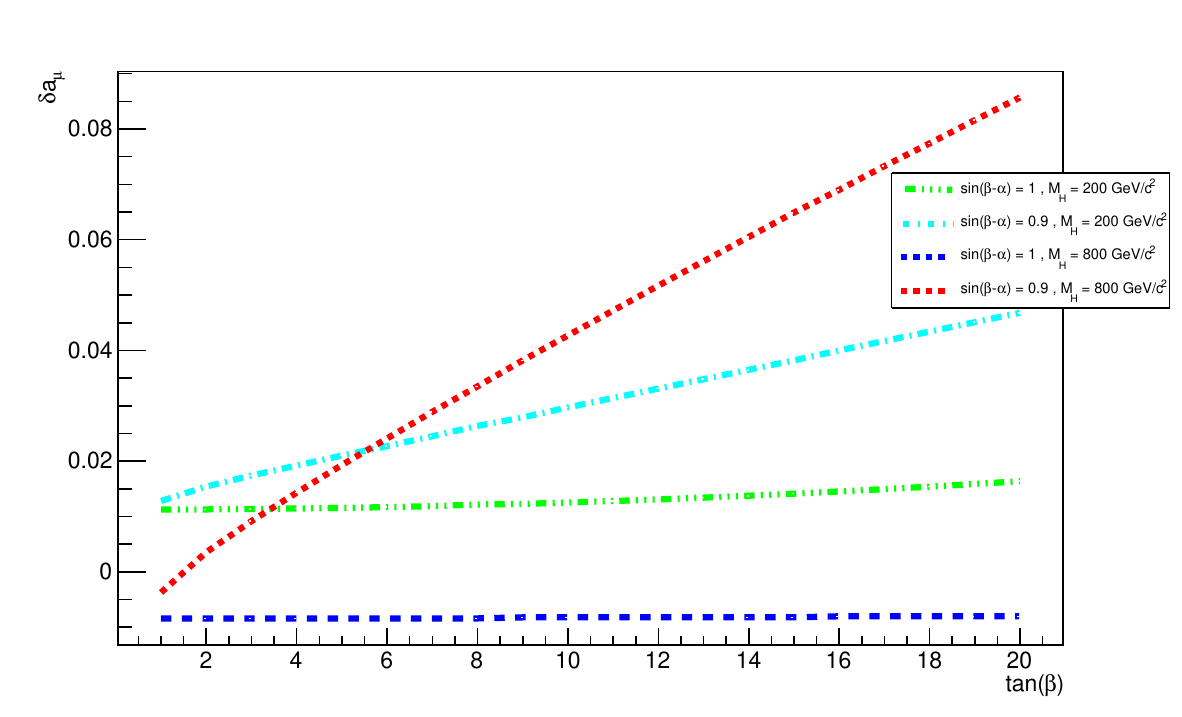}
	\caption{Deviation in muon anomalous magnetic moment ($\delta a_{\mu}$) versus  $\tan\beta$, in different conditions.}
	\label{figs:delta}
\end{figure}

\begin{figure}[!htb]
	\includegraphics*[width=.495\textwidth]{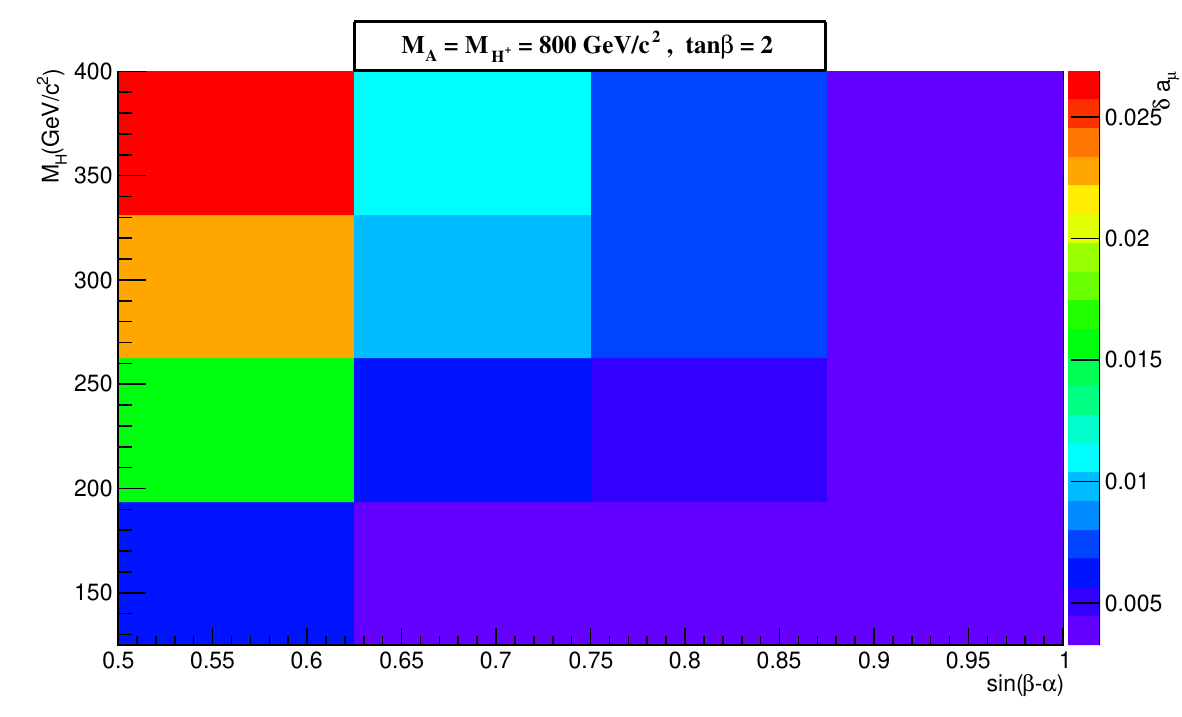}
	\includegraphics*[width=.495\textwidth]{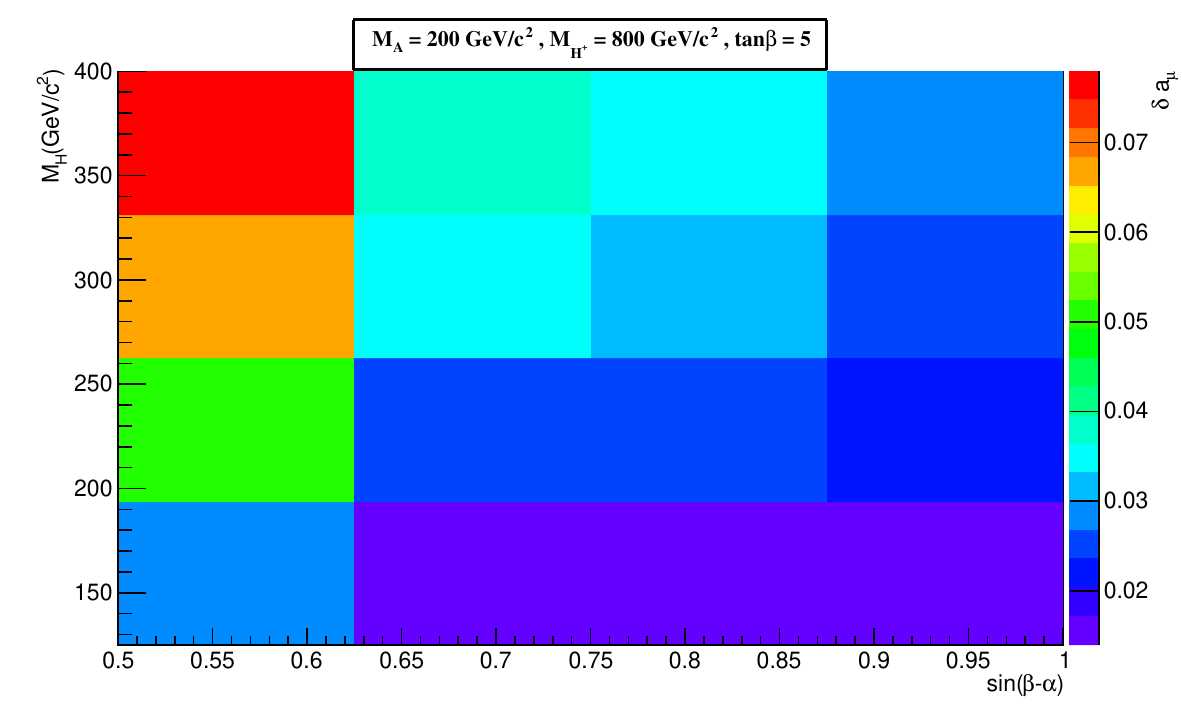}
	\includegraphics*[width=.495\textwidth]{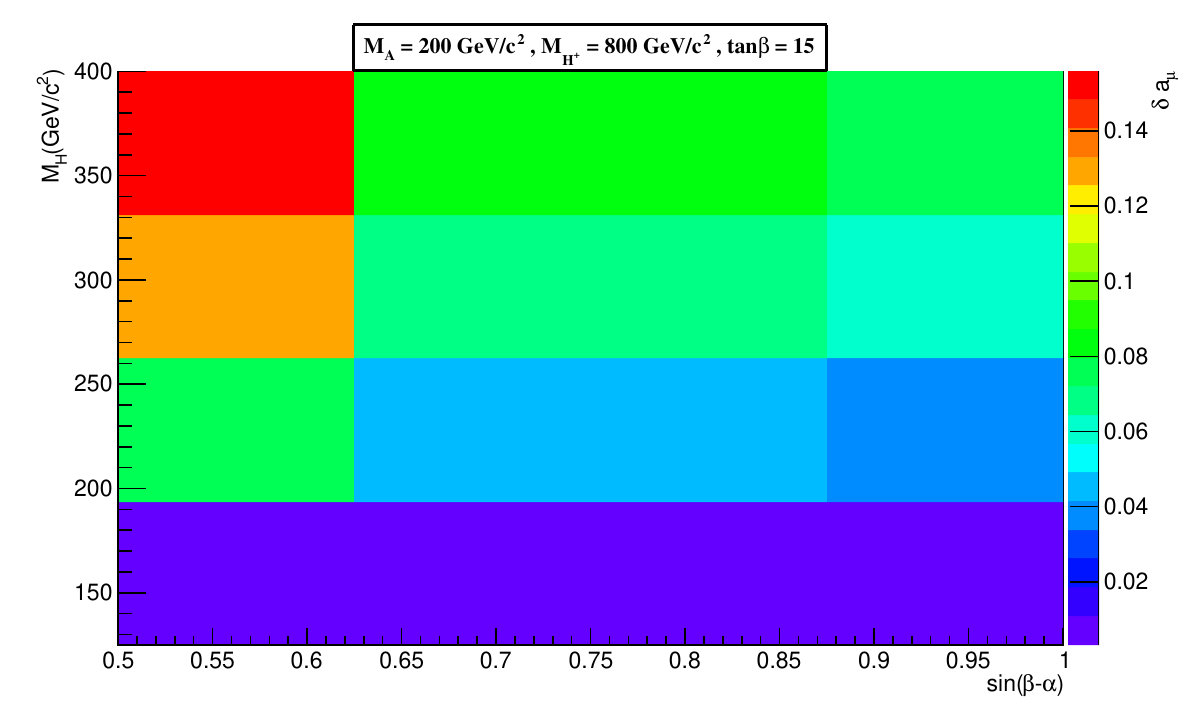}
	\includegraphics*[width=.495\textwidth]{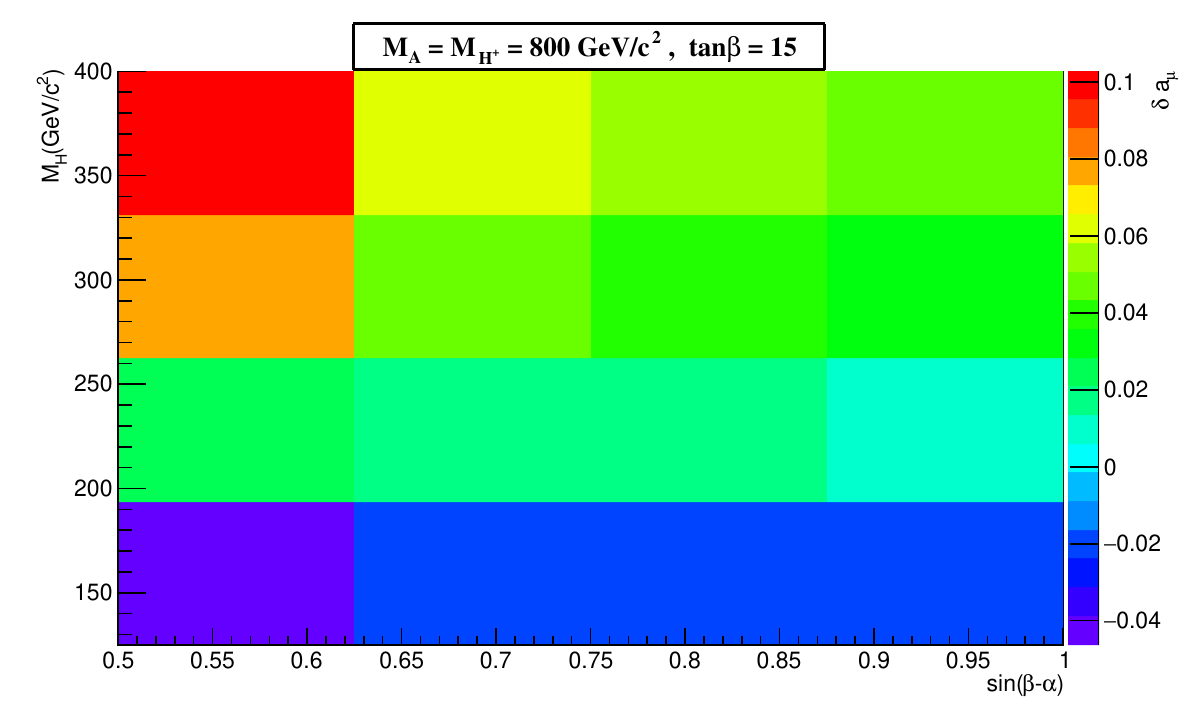}
	\caption{The distribution of $\delta a_{\mu}$ in the ($ M_{H} $, $\sin(\beta-\alpha) $) plane, in different conditions.}
	\label{SuperIso:AmuinPlane}
\end{figure}

In contrast to the BRBs, $\delta a_{\mu}$ depends on  $\sin(\beta-\alpha)$ strongly but, depends on $M_{H^{\pm}}$ and $M_A$ weakly.
Figure \ref{figs:delta} 
shows the behavior of  $\delta a_{\mu}$ versus $\tan\beta$ in different values of  $\sin(\beta-\alpha)$ and  $M_{H}$. To produce these curves, $M_{H^{\pm}}$ is set to 800 GeV/$c^2$ and A and H are degenerate in mass. The main dependence is on  $\sin(\beta-\alpha)$. At $1\sigma$ and $2\sigma$ precision, $\delta a_{\mu}$ does not lie in the consistent interval in the whole considered plane.
Therefore, even at $2\sigma$ precision there are not any points that satisfy all five observable measurements simultaneously. 
Figure \ref{SuperIso:AmuinPlane} 
shows the $\delta a_{\mu}$  distribution in the parameter space which is used in the cross section analysis ($M_{H}$ vs. $\sin(\beta-\alpha)$). Different shapes correspond to changing the other parameters values. Various values for $\tan\beta$, $M_{H^{\pm}}$, and $M_{A}$ are examined but, similar behaviour is seen. Usually, the largest value of $\delta a_{\mu}$ happens in low values of $\sin(\beta-\alpha)$ and high values of  $M_{H}$. The smallest values happen in low values of $M_{H}$ or high values of $\sin(\beta-\alpha)$, depending on the values of $\tan\beta$.

It should be noted that the direct search constraints are the results of the searches for 2HDM particles in different colliders. Inconsistency with these constraints means that the test point in question is excluded by some experiments. In contrary the flavour physics constraints check if existence of 2HDM in a special test point is consistent with some measurements. For example, $\delta a_{\mu}$ shows a large deviation between SM prediction and experimental measurement for $a_{\mu}$. Inconsistency between the reported test points and the experimental measurement shows that these special test points can not fill completely the gap between the prediction and experiment and may be some other new physics are needed to narrow the deviation.

\section{Conclusion}\label{sec:con} 
The measured cross section for Zh production has still large uncertainties. The idea of this research work is  to measure the cross section of Zh production at LHC
to find more bounds on the 2HDM type-II. 
To illustrate the idea, two scenarios are considered. In the first scenario: it is assumed that the measured value for the Zh production cross section is very close to the predicted value of the SM with an uncertainty of $10\%$, in the other words, the 2HDM light Higgs boson is  same as the SM Higgs boson. In this case, in the low masses of the heavy CP-even Higgs boson, there is not any restrictions on the values of $\sin(\beta-\alpha)$, while at the high values of $\sin(\beta-\alpha)$ there is not any limits on the heavy CP-even Higgs mass. 
In the second scenario, when 75\% of the SM predicted cross section is measured at LHC, the alignment limit is not anymore allowed. For low values of $\sin(\beta-\alpha)$, only the heavy CP-even Higgs masses close to 140 GeV/$c^2$ are allowed. The results of this analysis are combined with the previous results of the experimental searches using the HiggsBounds code. This combination can constrain the 2HDM phase space tightly. It is suggested to the LHC experiments to measure the Zh cross section more precisely to be able to decide about the existence of 2HDM. To complete the analysis, constraints from flavour physics are also studied using the SuperIso package. Although BR($B_{s}^{0} \longrightarrow \mu^{+} \mu^{-}$) measurements can prefer some parts of the parameter space but, other investigated observables can not add a special point to the results.

\section{Data Availability Statement} No Data associated in the manuscript

%


\begin{thebibliography}{26}
	\expandafter\ifx\csname natexlab\endcsname\relax\def\natexlab#1{#1}\fi
	\expandafter\ifx\csname bibnamefont\endcsname\relax
	\def\bibnamefont#1{#1}\fi
	\expandafter\ifx\csname bibfnamefont\endcsname\relax
	\def\bibfnamefont#1{#1}\fi
	\expandafter\ifx\csname citenamefont\endcsname\relax
	\def\citenamefont#1{#1}\fi
	\expandafter\ifx\csname url\endcsname\relax
	\def\url#1{\texttt{#1}}\fi
	\expandafter\ifx\csname urlprefix\endcsname\relax\def\urlprefix{URL }\fi
	\providecommand{\bibinfo}[2]{#2}
	\providecommand{\eprint}[2][]{\url{#2}}
	
	\bibitem[{\citenamefont{Aad et~al.}(2012)}]{ATLAS:2012yve}
	\bibinfo{author}{\bibfnamefont{G.}~\bibnamefont{Aad}} \bibnamefont{et~al.}
	(\bibinfo{collaboration}{ATLAS}), \bibinfo{journal}{Phys. Lett. B}
	\textbf{\bibinfo{volume}{716}}, \bibinfo{pages}{1} (\bibinfo{year}{2012}),
	\eprint{1207.7214}.
	
	\bibitem[{\citenamefont{Chatrchyan et~al.}(2012)}]{CMS:2012qbp}
	\bibinfo{author}{\bibfnamefont{S.}~\bibnamefont{Chatrchyan}}
	\bibnamefont{et~al.} (\bibinfo{collaboration}{CMS}), \bibinfo{journal}{Phys.
		Lett. B} \textbf{\bibinfo{volume}{716}}, \bibinfo{pages}{30}
	(\bibinfo{year}{2012}), \eprint{1207.7235}.
	
	\bibitem[{\citenamefont{Lykken}(2010)}]{Lykken:2010mc}
	\bibinfo{author}{\bibfnamefont{J.~D.} \bibnamefont{Lykken}}
	(\bibinfo{year}{2010}), \eprint{1005.1676}.
	
	\bibitem[{\citenamefont{Ma and Ng}(1994)}]{Ma_1994}
	\bibinfo{author}{\bibfnamefont{E.}~\bibnamefont{Ma}} \bibnamefont{and}
	\bibinfo{author}{\bibfnamefont{D.}~\bibnamefont{Ng}},
	\bibinfo{journal}{Physical Review D} \textbf{\bibinfo{volume}{49}},
	\bibinfo{pages}{6164} (\bibinfo{year}{1994}),
	\urlprefix\url{https://doi.org/10.1103/physrevd.49.6164}.
	
	\bibitem[{\citenamefont{Haber and Kane}(1985)}]{HABER198575}
	\bibinfo{author}{\bibfnamefont{H.}~\bibnamefont{Haber}} \bibnamefont{and}
	\bibinfo{author}{\bibfnamefont{G.}~\bibnamefont{Kane}},
	\bibinfo{journal}{Physics Reports} \textbf{\bibinfo{volume}{117}},
	\bibinfo{pages}{75} (\bibinfo{year}{1985}), ISSN \bibinfo{issn}{0370-1573},
	\urlprefix\url{https://www.sciencedirect.com/science/article/pii/0370157385900511}.
	
	\bibitem[{\citenamefont{Carena and Haber}(2003)}]{Carena_2003}
	\bibinfo{author}{\bibfnamefont{M.}~\bibnamefont{Carena}} \bibnamefont{and}
	\bibinfo{author}{\bibfnamefont{H.}~\bibnamefont{Haber}},
	\bibinfo{journal}{Progress in Particle and Nuclear Physics}
	\textbf{\bibinfo{volume}{50}}, \bibinfo{pages}{63} (\bibinfo{year}{2003}),
	\urlprefix\url{https://doi.org/10.1016/s0146-6410(02)00177-1}.
	
	\bibitem[{\citenamefont{Davidson and Haber}(2005)}]{PhysRevD.72.035004}
	\bibinfo{author}{\bibfnamefont{S.}~\bibnamefont{Davidson}} \bibnamefont{and}
	\bibinfo{author}{\bibfnamefont{H.~E.} \bibnamefont{Haber}},
	\bibinfo{journal}{Phys. Rev. D} \textbf{\bibinfo{volume}{72}},
	\bibinfo{pages}{035004} (\bibinfo{year}{2005}),
	\urlprefix\url{https://link.aps.org/doi/10.1103/PhysRevD.72.035004}.
	
	\bibitem[{\citenamefont{DJOUADI}(2008)}]{DJOUADI_2008}
	\bibinfo{author}{\bibfnamefont{A.}~\bibnamefont{Djouadi}},
	\bibinfo{journal}{Physics Reports} \textbf{\bibinfo{volume}{459}},
	\bibinfo{pages}{1} (\bibinfo{year}{2008}),
	\urlprefix\url{https://doi.org/10.1016/j.physrep.2007.10.005}.
	
	\bibitem[{\citenamefont{Branco et~al.}(2012)\citenamefont{Branco, Ferreira,
			Lavoura, Rebelo, Sher, and Silva}}]{Branco:2011iw}
	\bibinfo{author}{\bibfnamefont{G.~C.} \bibnamefont{Branco}},
	\bibinfo{author}{\bibfnamefont{P.~M.} \bibnamefont{Ferreira}},
	\bibinfo{author}{\bibfnamefont{L.}~\bibnamefont{Lavoura}},
	\bibinfo{author}{\bibfnamefont{M.~N.} \bibnamefont{Rebelo}},
	\bibinfo{author}{\bibfnamefont{M.}~\bibnamefont{Sher}}, \bibnamefont{and}
	\bibinfo{author}{\bibfnamefont{J.~P.} \bibnamefont{Silva}},
	\bibinfo{journal}{Phys. Rept.} \textbf{\bibinfo{volume}{516}},
	\bibinfo{pages}{1} (\bibinfo{year}{2012}), \eprint{1106.0034}.
	
	\bibitem[{\citenamefont{Ayazi and Paktinat}(2019)}]{Ayazi:2019kli}
	\bibinfo{author}{\bibfnamefont{S.~Y.} \bibnamefont{Ayazi}} \bibnamefont{and}
	\bibinfo{author}{\bibfnamefont{S.}~\bibnamefont{Paktinat}},
	\bibinfo{journal}{J. Exp. Theor. Phys.} \textbf{\bibinfo{volume}{128}},
	\bibinfo{pages}{865} (\bibinfo{year}{2019}), \eprint{1707.07155}.
	
	\bibitem[{\citenamefont{Tumasyan et~al.}(2023)}]{CMS:2023vzh}
	\bibinfo{author}{\bibfnamefont{A.}~\bibnamefont{Tumasyan}} \bibnamefont{et~al.}
	(\bibinfo{collaboration}{CMS}) (\bibinfo{year}{2023}), \eprint{2312.07562}.
	
	\bibitem[{\citenamefont{Aad et~al.}(2023)}]{ATLAS:2023qpu}
	\bibinfo{author}{\bibfnamefont{G.}~\bibnamefont{Aad}} \bibnamefont{et~al.}
	(\bibinfo{collaboration}{ATLAS}) (\bibinfo{year}{2023}), \eprint{2312.02394}.
	
	\bibitem[{\citenamefont{Haber and O'Neil}(2006)}]{PhysRevD.74.015018}
	\bibinfo{author}{\bibfnamefont{H.~E.} \bibnamefont{Haber}} \bibnamefont{and}
	\bibinfo{author}{\bibfnamefont{D.}~\bibnamefont{O'Neil}},
	\bibinfo{journal}{Phys. Rev. D} \textbf{\bibinfo{volume}{74}},
	\bibinfo{pages}{015018} (\bibinfo{year}{2006}),
	\urlprefix\url{https://link.aps.org/doi/10.1103/PhysRevD.74.015018}.
	
	\bibitem[{\citenamefont{Eberhardt et~al.}(2013)\citenamefont{Eberhardt,
			Nierste, and Wiebusch}}]{Eberhardt:2013uba}
	\bibinfo{author}{\bibfnamefont{O.}~\bibnamefont{Eberhardt}},
	\bibinfo{author}{\bibfnamefont{U.}~\bibnamefont{Nierste}}, \bibnamefont{and}
	\bibinfo{author}{\bibfnamefont{M.}~\bibnamefont{Wiebusch}},
	\bibinfo{journal}{JHEP} \textbf{\bibinfo{volume}{07}}, \bibinfo{pages}{118}
	(\bibinfo{year}{2013}), \eprint{1305.1649}.
	
	\bibitem[{\citenamefont{Ginzburg and Krawczyk}(2005)}]{Ginzburg_2005}
	\bibinfo{author}{\bibfnamefont{I.~F.} \bibnamefont{Ginzburg}} \bibnamefont{and}
	\bibinfo{author}{\bibfnamefont{M.}~\bibnamefont{Krawczyk}},
	\bibinfo{journal}{Physical Review D} \textbf{\bibinfo{volume}{72}}
	(\bibinfo{year}{2005}),
	\urlprefix\url{https://doi.org/10.1103/physrevd.72.115013}.
	
	\bibitem[{\citenamefont{Eriksson et~al.}(2010)\citenamefont{Eriksson, Rathsman,
			and Stal}}]{Eriksson:2009ws}
	\bibinfo{author}{\bibfnamefont{D.}~\bibnamefont{Eriksson}},
	\bibinfo{author}{\bibfnamefont{J.}~\bibnamefont{Rathsman}}, \bibnamefont{and}
	\bibinfo{author}{\bibfnamefont{O.}~\bibnamefont{Stal}},
	\bibinfo{journal}{Comput. Phys. Commun.} \textbf{\bibinfo{volume}{181}},
	\bibinfo{pages}{189} (\bibinfo{year}{2010}), \eprint{0902.0851}.
	
	\bibitem[{\citenamefont{Alwall et~al.}(2011)\citenamefont{Alwall, Herquet,
			Maltoni, Mattelaer, and Stelzer}}]{Alwall:2011uj}
	\bibinfo{author}{\bibfnamefont{J.}~\bibnamefont{Alwall}},
	\bibinfo{author}{\bibfnamefont{M.}~\bibnamefont{Herquet}},
	\bibinfo{author}{\bibfnamefont{F.}~\bibnamefont{Maltoni}},
	\bibinfo{author}{\bibfnamefont{O.}~\bibnamefont{Mattelaer}},
	\bibnamefont{and} \bibinfo{author}{\bibfnamefont{T.}~\bibnamefont{Stelzer}},
	\bibinfo{journal}{JHEP} \textbf{\bibinfo{volume}{06}}, \bibinfo{pages}{128}
	(\bibinfo{year}{2011}), \eprint{1106.0522}.
	
	\bibitem[{\citenamefont{Brun and Rademakers}(1997)}]{Brun:1997pa}
	\bibinfo{author}{\bibfnamefont{R.}~\bibnamefont{Brun}} \bibnamefont{and}
	\bibinfo{author}{\bibfnamefont{F.}~\bibnamefont{Rademakers}},
	\bibinfo{journal}{Nucl. Instrum. Meth. A} \textbf{\bibinfo{volume}{389}},
	\bibinfo{pages}{81} (\bibinfo{year}{1997}).
	
	\bibitem[{\citenamefont{Bini}(2014)}]{Bini:2014yhn}
	\bibinfo{author}{\bibfnamefont{C.}~\bibnamefont{Bini}} (\bibinfo{year}{2014}),
	\urlprefix\url{https://www.roma1.infn.it/~bini/StatEPP_new.pdf}.
	
	\bibitem[{\citenamefont{Kanemura et~al.}(1993)\citenamefont{Kanemura, Kubota,
			and Takasugi}}]{KANEMURA1993155}
	\bibinfo{author}{\bibfnamefont{S.}~\bibnamefont{Kanemura}},
	\bibinfo{author}{\bibfnamefont{T.}~\bibnamefont{Kubota}}, \bibnamefont{and}
	\bibinfo{author}{\bibfnamefont{E.}~\bibnamefont{Takasugi}},
	\bibinfo{journal}{Physics Letters B} \textbf{\bibinfo{volume}{313}},
	\bibinfo{pages}{155} (\bibinfo{year}{1993}), ISSN \bibinfo{issn}{0370-2693},
	\urlprefix\url{https://www.sciencedirect.com/science/article/pii/0370269393912052}.
	
	\bibitem[{\citenamefont{Gunion et~al.}(2000)\citenamefont{Gunion, Haber, Kane,
			and Dawson}}]{Gunion:1989we}
	\bibinfo{author}{\bibfnamefont{J.~F.} \bibnamefont{Gunion}},
	\bibinfo{author}{\bibfnamefont{H.~E.} \bibnamefont{Haber}},
	\bibinfo{author}{\bibfnamefont{G.~L.} \bibnamefont{Kane}}, \bibnamefont{and}
	\bibinfo{author}{\bibfnamefont{S.}~\bibnamefont{Dawson}},
	\emph{\bibinfo{title}{{The Higgs Hunter's Guide}}}, vol.~\bibinfo{volume}{80}
	(\bibinfo{year}{2000}).
	
	
	\bibitem[{\citenamefont{Degrande}(2015)}]{Degrande_2015}
	\bibinfo{author}{\bibfnamefont{C.}~\bibnamefont{Degrande}},
	\bibinfo{journal}{Computer Physics Communications}
	\textbf{\bibinfo{volume}{197}}, \bibinfo{pages}{239} (\bibinfo{year}{2015}),
	\urlprefix\url{https://doi.org/10.1016/j.cpc.2015.08.015}.
	
	\bibitem[{\citenamefont{Denner et~al.}(2018)\citenamefont{Denner, Lang, and
			Uccirati}}]{Denner_2018}
	\bibinfo{author}{\bibfnamefont{A.}~\bibnamefont{Denner}},
	\bibinfo{author}{\bibfnamefont{J.-N.} \bibnamefont{Lang}}, \bibnamefont{and}
	\bibinfo{author}{\bibfnamefont{S.}~\bibnamefont{Uccirati}},
	\bibinfo{journal}{Computer Physics Communications}
	\textbf{\bibinfo{volume}{224}}, \bibinfo{pages}{346} (\bibinfo{year}{2018}),
	\urlprefix\url{https://doi.org/10.1016/j.cpc.2017.11.013}.
	
	\bibitem[{\citenamefont{Bechtle et~al.}(2020)\citenamefont{Bechtle, Dercks,
			Heinemeyer, Klingl, Stefaniak, Weiglein, and Wittbrodt}}]{Bechtle:2020pkv}
	\bibinfo{author}{\bibfnamefont{P.}~\bibnamefont{Bechtle}},
	\bibinfo{author}{\bibfnamefont{D.}~\bibnamefont{Dercks}},
	\bibinfo{author}{\bibfnamefont{S.}~\bibnamefont{Heinemeyer}},
	\bibinfo{author}{\bibfnamefont{T.}~\bibnamefont{Klingl}},
	\bibinfo{author}{\bibfnamefont{T.}~\bibnamefont{Stefaniak}},
	\bibinfo{author}{\bibfnamefont{G.}~\bibnamefont{Weiglein}}, \bibnamefont{and}
	\bibinfo{author}{\bibfnamefont{J.}~\bibnamefont{Wittbrodt}},
	\bibinfo{journal}{Eur. Phys. J. C} \textbf{\bibinfo{volume}{80}},
	\bibinfo{pages}{1211} (\bibinfo{year}{2020}), \eprint{2006.06007}.
	
	\bibitem[{\citenamefont{Bifani et~al.}(2019)\citenamefont{Bifani,
			Descotes-Genon, Romero~Vidal, and Schune}}]{Bifani:2018zmi}
	\bibinfo{author}{\bibfnamefont{S.}~\bibnamefont{Bifani}},
	\bibinfo{author}{\bibfnamefont{S.}~\bibnamefont{Descotes-Genon}},
	\bibinfo{author}{\bibfnamefont{A.}~\bibnamefont{Romero~Vidal}},
	\bibnamefont{and} \bibinfo{author}{\bibfnamefont{M.-H.}
		\bibnamefont{Schune}}, \bibinfo{journal}{J. Phys. G}
	\textbf{\bibinfo{volume}{46}}, \bibinfo{pages}{023001}
	(\bibinfo{year}{2019}), \eprint{1809.06229}.
	
	\bibitem[{\citenamefont{Workman et~al.}(2022)}]{ParticleDataGroup:2022pth}
	\bibinfo{author}{\bibfnamefont{R.~L.} \bibnamefont{Workman}}
	\bibnamefont{et~al.} (\bibinfo{collaboration}{Particle Data Group}),
	\bibinfo{journal}{PTEP} \textbf{\bibinfo{volume}{2022}},
	\bibinfo{pages}{083C01} (\bibinfo{year}{2022}).
	
	\bibitem[{\citenamefont{Neshatpour and Mahmoudi}(2022)}]{Neshatpour:2022fak}
	\bibinfo{author}{\bibfnamefont{S.}~\bibnamefont{Neshatpour}} \bibnamefont{and}
	\bibinfo{author}{\bibfnamefont{F.}~\bibnamefont{Mahmoudi}},
	\bibinfo{journal}{PoS} \textbf{\bibinfo{volume}{CompTools2021}},
	\bibinfo{pages}{010} (\bibinfo{year}{2022}), \eprint{2207.04956}.
	
	\bibitem[{\citenamefont{Mahmoudi}(2009)}]{Mahmoudi:2008tp}
	\bibinfo{author}{\bibfnamefont{F.}~\bibnamefont{Mahmoudi}},
	\bibinfo{journal}{Comput. Phys. Commun.} \textbf{\bibinfo{volume}{180}},
	\bibinfo{pages}{1579} (\bibinfo{year}{2009}), \eprint{0808.3144}.
	
\end{thebibliography}
\end{document}